
\hoffset .5 true in
\hsize = 6.0  true in
\vsize = 9.0 true in
\baselineskip = 15pt plus 2pt   
\parskip = 6pt			
\font\tentworm=cmr10 scaled \magstep2
\font\tentwobf=cmbx10 scaled \magstep2
\font\tenonerm=cmr10 scaled \magstep1
\font\tenonebf=cmbx10 scaled \magstep1
\font\eightrm=cmr8
\font\eightit=cmti8
\font\eightbf=cmbx8
\font\eightsl=cmsl8
\font\sevensy=cmsy7
\font\sevenm=cmmi7
\font\twelverm=cmr12
\font\twelvebf=cmbx12

\def\subsection #1\par{\noindent {\bf #1} \noindent \rm}
\def\mid {\let\rm=\tenonerm \let\bf=\tenonebf \rm \bf}
\def\para{\par \vskip 12 pt}
\def\head{\let\rm=\tentworm \let\bf=\tentwobf \rm \bf}
\def\heading #1 #2\par{\centerline {\head #1} \smallskip
 \centerline {\head #2} \vskip .15 pt \rm}
\def\eight{\let\rm=\eightrm \let\it=\eightit \let\bf=\eightbf
\let\sl=\eightsl \let\sy=\sevensy \let\m=\sevenm \rm}

\def\foots{\noindent \eight \baselineskip=10 true pt \noindent \rm}
\def\sexion{\let\rm=\twelverm \let\bf=\twelvebf \rm \bf}

\def\section #1 #2\par{\vskip 20 pt \noindent {\mid #1} \enspace {\mid #2}
  \para \noindent \rm}

\def\ssection #1 #2\par{\noindent {\mid #1} \enspace {\mid #2}
  \para \noindent \rm}

\def\abstract#1\par{\para \foots {\bf Abstract: \enspace}#1 \para}

\def\author#1\par{\centerline {#1} \vskip 0.1 true in \rm}

\def\abstract#1\par{\noindent {\bf Abstract: }#1 \vskip 0.5 true in \rm}

\def\midsection #1\par{\noindent {\sexion #1} \noindent \rm}

\def\sqr#1#2{{\vcenter{\vbox{\hrule height#2pt
 \hbox {\vrule width#2pt height#1pt \kern#1pt
  \vrule width#2pt}
  \hrule height#2pt}}}}

\def \dk {|\delta_k|}
\def \dm {{\Delta M\over M}}
\def \der {\bigg({\delta\rho\over \rho}\bigg)_H}
\def \dr2 {\bigg({\delta\rho\over \rho}\bigg)_H^2}

\def \hmp2 {\bigg({H_1\over m_{\rm p}}\bigg)^2}

\def \e {\eta }
\def \ee {({\eta\over \eta _0})}
\def \g {\Gamma}

\def \m {|\mu|}

\def \half {1\over 2}

\def \jetp {Sov. Phys. JETP}
\def \jetpl {JETP Lett.}

\def \ap {Astrophys. J}
\def \pd {Phys. Rev. D}
\def \prl {Phys. Rev. Lett.}
\def \pl {Phys. Lett.}
\def \np {Nucl. Phys.}

\def \doublespace {\baselineskip = 20pt plus 7pt \message {double space}}
\def \singlespace {\baselineskip = 13pt plus 3pt \message {single space}}
\singlespace

\def \body {\vfill \eject \parindent = 1.0 true cm
	\ifx \spacing \singlespace \singlespace \else \doublespace \fi}

\def \title#1 {\centerline {{\bf #1}}}
\def \Abstract#1 {\noindent \baselineskip=15pt plus 3pt \parshape=1 40pt310pt
  {\bf Abstract} \ \ #1}





\catcode`@=11
\def \C@ncel#1#2 {\ooalign {$\hfil#1 \mkern2mu/ \hfil $\crcr$#1#2$}}
\def \gf#1 {\mathrel {\mathpalette \c@ncel#1}}	
\def \Gf#1 {\mathrel {\mathpalette \C@ncel#1}}	

\def \gapx {\;\lower 2pt \hbox {$\buildrel > \over {\scriptstyle {\sim}}$}
\; }
\def \lapx {\;\lower 2pt \hbox {$\buildrel < \over {\scriptstyle {\sim}}$}
\; }


\topskip 0.7 true cm

\footline = {\ifnum \pageno = 1 \hfill \else \hfill \number \pageno \hfill \fi}

\def\half{{1\over 2}}
\def\n{\noindent}

\def\v2{\vskip 0.2 true cm}
\def\v3{\vskip 0.3 true cm}

\line{\hfill \foots{\bf IUCAA Preprint}}
\line{\hfill \foots{July, 1992}}

\vskip .45 true in

\heading {Density Perturbations, Gravity Waves  }

\heading { and  the }

\heading { Cosmic Microwave Background}

\vfill
\centerline{by}
\smallskip
\centerline{Tarun Souradeep$^*$ and Varun Sahni$^{\dag}$}
\medskip
\centerline{Inter-University Centre for Astronomy and Astrophysics}
\centerline{Post Bag 4, Ganeshkhind, Pune 411 007}
\centerline{INDIA}

\vfill
\settabs 10\columns
\+&&&email~:&$^*$~~~~tarun@iucaa.ernet.in \cr
\+&&&&$\dag$~~~~varun@iucaa.ernet.in\cr
\medskip

\vfil\eject


\section { Abstract}

We asses the contribution to the observed large scale anisotropy of the cosmic
microwave background radiation, arising from both gravity waves as well as
adiabatic density perturbations, generated by a common inflationary mechanism
in the early Universe.
 We find that for inflationary models predicting power law primordial spectra
$\dk^2 \propto k^n$, the relative contribution to the quadrupole
anisotropy from gravity waves and scalar density perturbations, depends
crucially upon $n$. For $n < 0.87$, gravity waves perturb the CMBR by a larger
amount than density perturbations, whereas for $n > 0.87$ the reverse is true.
Normalising the amplitude of the density perturbation spectrum at large scales,
using the observed value of the COBE quadrupole, we determine
$(\delta M /M)_{16}$ -- the rms density contrast on scales
$\sim 16 h_{50}^{-1}$
Mpc, for cosmological models with cold dark matter. We find that for
$n < 0.75$, a large amount of biasing is required in order to reconcile theory
with observations.

\bigskip
\bigskip

\item {} {\bf keywords :} structure formation, density fluctuations, biased
galaxy
formation, gravity waves, inflation, microwave anisotropy, CMBR

\vskip 1.0cm
\vfill\eject
\topskip  0.7cm
\bigskip


Both gravity waves as well as adiabatic density perturbations arise as natural
consequences of the inflationary scenario, due to the superadiabatic
amplification of zero-point quantum fluctuations occuring during inflation
(~Starobinsky 1979, 1982; Hawking 1982; Guth \& Pi 1982~).
As gravity waves and scalar density perturbations enter the horizon
during matter domination, they induce distortions in the cosmic microwave
background radiation (CMBR) through the Sach-Wolfe effect. In this paper we
shall assess the relative contribution to the CMBR anisotropy, arising from
both
scalar density perturbations as well as from gravity waves.
In doing so we shall consider waves created
during exponential ($a \propto
\exp {Ht}$), quasi-exponential ($a \propto {\rm exp}{\int H(t)dt}$),
and power law ($a \propto t^p, p> 1$) inflation, thus
considerably extending previous work on the subject.

The close similarity between gravity waves
and massless scalars, in a Friedman - Robertson - Walker (FRW) background
(Grishchuk 1974),
makes it possible for each polarisation state of the graviton to be expressed
in terms of solutions to the massless, minimally coupled Klein-Gordon
equation: $h_{\times(+)} = \sqrt{8\pi G}~(\chi_k /a(\e))  \, {\rm e}^{-i{ \bf
k}.{\bf x}}\,$

$$
\ddot \chi_k + [k^2 - {\ddot a\over a}]\, \chi_k = 0
\eqno(1a)
$$

\noindent where $a(\e)$ is the scale factor of the Universe:
$ ds^2 = a^2 (\e)(d\e^2 - d{\bf x }\, ^2)$,
and $k$ is the comoving wavenumber $k = 2\pi a/\lambda$.
Equation (1a) closely resembles the Schroedinger equation in quantum mechanics,
with ${\ddot a / a}$, playing the role of the potential barrier `V'.
(The form of $V \equiv (\ddot a/ a)$, is shown in Figure 1, for inflation
followed by a matter dominated epoch.) Solutions of equation (1a) have some
interesting properties, for instance, small wavelength solutions of (1a)
are adiabatically damped:

$$\phi_k ^ + (\e) \equiv {\chi_k\over a(\e)} ~~{\mathop{=}_{k\e \gg 2\pi
} } {1 \over \sqrt{ 2k} a(\e) }~{\exp (-ik\e)}, \eqno(1b)$$
whereas long wavelength modes asymptotically approach the form :
$$
\phi_k (\e) \equiv {\chi_k\over a(\e)} ~~ {\mathrel{\mathop{=}_{k\e \ll 2\pi} }
}
 A(k) + B(k) \int {d\ee \over a^2}~,
 \eqno(1c)
 $$
 \noindent where A and B are constants. The form of (1c) implies that the
amplitude of a fluctuation
 (equivalently -- gravity wave), freezes to a constant value when its
 wavelength (during inflation), becomes larger than the corresponding
 Hubble radius. Since modes are continuously being pushed outside the
 Hubble radius during inflation, a mode with a fixed wavenumber ``k", which
left
 the Hubble radius during inflation ($k = 2 \pi ~ \eta_1^{-1}$),
 will re-enter it later
 during the epoch of matter domination ($k = 2 \pi ~\eta_0^{-1}$).
 The amplitude
 of the mode between these two epochs is super-adiabatically amplified,
 which in field-theoretic language corresponds to the quantum creation of
 gravitons (Grishchuk 1974). Graviton production in an inflationary Universe
 has been studied in
 some detail by Starobinsky (1979), Abbott \& Wise (1984), Allen (1988), and
 Sahni (1990).
 In particular, was demonstrated that the
amplitude of gravity waves on scales greater than
the Hubble radius, has the time independent form (~Sahni 1990~)

$$ h_{+,\times}  = \sqrt{16 \pi G}
{k^{3 \over 2}\over \sqrt 2 \pi} |\phi_k|\Big|_{k|\e _1| = 2\pi}
= \sqrt{16 \pi} A(\nu) ~\bigg({H_1\over m_{\rm p}}\bigg),~~~~A(\nu) =
\g (\nu)\pi^{-\nu}(\nu - \half)^{-1}, \eqno(2a)$$

\noindent where we have summed over both polarisation states of the graviton.
$H_1$ is the value of the Hubble parameter at a time $k|\e _1| = 2\pi$,
when the given mode left the Hubble radius.
For exponential inflation
$\nu = {3\over 2}$ and $h_{+,\times} = (2/\sqrt{\pi})
{}~ (H_1/ m_{\rm p})$, in agreement with Abbott and Wise (1984).
($m_{\rm p}$ is the planck mass, $m_{\rm p} = 1.2 \times 10^{19} GeV$.)

Long wavelength gravity waves, entering the Horizon today, perturb the
cosmic microwave background through the Sachs - Wolfe effect
(~Sachs \& Wolfe 1967; Rubakov, Sazhin \& Veryaskin 1982;
Fabbri \& Pollock 1983; Abbott \& Wise 1984~). Writing
$(\delta T /T)$ in terms of a multipole expansion:

$${\delta T \over T} = \sum_{l,m}{\rm a}_{lm} Y_l^m(\alpha,\delta), \eqno(3)$$

\noindent it can be shown that (~Abbott and Wise 1984~)

$$ {\rm a}_2^2 \equiv \langle|{\rm a}_{2m}|^2\rangle = 0.145~
h_{+,\times}^2 .\eqno(4a)$$

\noindent The corresponding value of the {\it rms} quadrupole amplitude is
$$
Q_{T}^2 = {5\over 4 \pi} ~{\rm a}_2^2 = 2.9~ A^2(\nu)~\bigg({H_1\over m_{\rm
p}}\bigg)^2.
\eqno(4b)$$

\noindent (For exponential inflation $\nu = {3\over 2}$, and
 $Q_{T}^2 = (2.9 / 4 \pi^2) (H_1/ m_{\rm p})^2.$)
(The subscript ``T", denotes the fact that the quadrupole anisotropy is induced
by {\it tensor} waves, gravity waves being quadrupolar, do not contribute to a
dipole component ${\rm a}_1$.)

In addition to gravity waves, scalar density perturbations generated during
inflation, also perturb the CMBR via the Sachs-Wolfe effect, caused by
potential fluctuations at the surface of last scattering,
$(\delta T/ T) \sim {1\over 3}~ (\delta \varphi /c^2)$.
Inflationary density perturbations are characterised by a spectrum
$\dk^2 = A~ k^n$, the scale invariant spectrum
$n = 1$, being predicted by inflationary models
with exponential expansion.  Models with power law inflation
($a(t) \propto t^p, p > 1$), which arise in a number of theories
including extended inflation (La \& Steinhardt 1989), predict a more
general value for the spectral index , $ n = (p - 3)/ (p - 1)~~~$
(Lucchin and Matarrese 1985a,b).

The amplitude of density perturbations at horizon crossing ($k = H_0/c$)
is given by (~Kolb and Turner 1990~)
$$
\dr2 \equiv {k^3 \dk^2 \over 2 \pi^2} = {A \over 2 \pi^2} H_0^{n+3}.
\eqno(5)
$$
(we assume $c = 1$, for simplicity.)
The associated fluctuations in the microwave background are given by
$$
{\rm a}_l^2 \equiv \langle|{\rm a}_{lm}|^2\rangle = {H_0^4\over 2\pi}
\int_0^\infty {dk\over k^2}\dk^2 j_l^2(kx)
\eqno(6a)
$$
where $x = (2/ H_0)$, is the present day horizon size. For power law
spectra $\dk^2 = A~k^n$, (6a) translates into (Gradshteyn \& Ryzhik 1980)
$$
{\rm a}_2^2 = {A H_0^{n+3}\over 16}~f(n) = {\pi^2\over 8} f(n) \dr2 ,
{}~~~~~ f(n) = {\Gamma(3 - n) \Gamma ({3+n\over 2})\over
\Gamma^2({4-n\over 2}) \Gamma({9-n\over 2})}.
\eqno(6b)
$$

\noindent Values of $n$ in the range $0.5 \le n \le 1.7$, are consistent with
the recent
COBE results (Smoot et al. 1992).
For $n = 1$, (6b) reduces to
${\rm a}_2^2 = (\pi/12)~(\delta\rho/\rho)^2_H $.

\noindent The related value of the {\it rms} quadrupole amplitude is given by
$$
Q_{S}^2 = {5\over 4\pi} {\rm a}_2^2 = {5\pi \over 32} f(n) \dr2
\eqno(7)
$$
(the subscript ``S" refers to {\it scalar} perturbations.)

As shown by a number of authors, scalar field fluctuations which left the
Hubble radius during inflation, upon re-entering the horizon during
radiation (matter) domination, give rise to a density fluctuation, whose
{\it rms} value is given by (Bardeen, Steinhardt \& Turner 1983)
$$
\der = b~ {H_1\over \dot \phi} ~\delta \phi
\eqno(8)
$$
where $H_1$ is the Hubble parameter at a time $t_1$, when a scale which
reenters the horizon at $t_0$, left the Hubble radius during the inflationary
epoch (see Figure 1).  For modes entering the horizon during radiation (matter)
 domination, $b = 4 ({2\over 5})$. (The main contribution to the observed
quadrupole anisotropy however, comes from modes entering the horizon during
matter domination, we shall therefore assume $b = {2\over 5}$ in the
ensuing discussion.) $\delta \phi$ arises because of quantum fluctuations
in the scalar field during inflation. For modes entering the horizon today
$\delta \phi$ is given by $\delta \phi = A(\nu) H_1$ (see (2)), from which we
recover the standard result $\delta \phi = {H_1 \over 2\pi}$, for exponential
inflation (~Bunch \& Davies 1978; Starobinsky 1982; Linde 1982 and
 Vilenkin \& Ford 1982~).

For power law inflation, the inflaton potential has the form
$ V(\phi) = \break V_0 \exp(- \sqrt{16\pi/ p}~(\phi/m_{\rm p}))$
which results in the following exact solution (~Burd \& Barrow 1988~) to the
field
equations: $H = (p/t)$, $(H/\dot\phi) = (\sqrt{4\pi p}/ m_{\rm p})$
,
as a result
$$
\der =   A(\nu)~{\sqrt{16\pi p}\over 5}~ {H_1\over m_{\rm p}}.
\eqno(9)
$$

For models with quasi-exponential inflation ($a \propto \exp{\int H(t)dt}$),
the slow-roll condition $\ddot \phi \simeq 0$, when incorporated into the
equation
of motion of the inflaton field $\ddot\phi + 3H\dot\phi + V'(\phi) = 0$,
allows (8) to be recast as
$$
\der = {16\over 5} \sqrt{{2\pi\over 3}} {V^{3\over 2}\over m_{\rm p}^3 |V'|}
\eqno(10)
$$
where $V(\phi)$ is the inflaton potential.
As the inflaton field {\it slow-rolls} down its potential, the Universe expands
quasi-exponentially $a \propto \exp {\int H(t)dt}$, where $\int H(t)dt \simeq
H(t) \Delta t$ must be greater than or equal to 64, in order to explain the
observed spatial flatness of the Universe.
Taking this into account we get (~Hawking 1985; Olive 1990~)
$$
{V'\over V} m_{\rm p} = ({\pi \over 8 \bar n})^\half,
\eqno(11)
$$
where $\bar n = (\bar N/64)$, and, $\bar N = H \Delta t$ is the approximate
number of e-foldings. Substituting (11) in (10) we finally obtain
$$
\der = {\sqrt{\bar n}\over 0.39} {H_1\over m_{\rm p}}.
\eqno(12)
$$
Both (9) as well as (12) express the density perturbation at horizon crossing
in terms of the Hubble parameter during inflation $H_1$.
Comparing (9) and (11) we find that for large values of $p ~~(p >> 1),$ an
interesting relationship exists between $\bar N$ and $p$, namely
$\bar N \simeq (p / 2)$. This result can be established in a
straightforward
manner by noting that, for slow-roll (quasi-exponential) inflation, the
number of e-foldings is given by
$$
N = \int_{\tau_1}^{\tau_2}H dt =
{8\pi \over m_{\rm p}^2}\int_{V_1}^{V_2} \bigg({V\over V'}\bigg)^2~{dV\over V}
\eqno(13a)
$$
(we have used the slow-roll condition $\dot\phi = - (V'/3H) <<
V^{1 \over 2}(\phi)$ as well as the Einstein equation $3H^2 \simeq
(8\pi/ m_{\rm p}^2) V$ in establishing (13a).~)
Substituting (11) in (13a) we finally obtain
$$
N = \bar N~ {\rm ln}~\bigg({V_1\over V_2}\bigg)
\eqno(13b)
$$
relating the approximate number of e-foldings $\bar N$, with the exact number
-- $N$.
We can establish a similar relationship for power law inflation by noting that
since $(H/ \dot\phi) = (\sqrt{4\pi p}/m_{\rm p})$ (~Burd \& Barrow 1988~),
we get
$$
N = \int_{\tau_1}^{\tau_2}H dt = {\sqrt {4\pi p}\over m_{\rm p}}
\int_{\tau_1}^{\tau_2}
\dot\phi dt = {\sqrt {4\pi p}\over m_{\rm p}} (\phi_2 - \phi_1) = {p\over 2}
{}~{\rm ln}~\bigg({V_1\over V_2}\bigg)\eqno(14)
$$
Comparing (13b) and (14) we find $\bar N = (p/2)$, which is what we set
out to establish.

{}From (7), (9) and (12) we finally obtain
$$
Q_{S}^2 = {5\over 48} \dr2 = {5\over 48}{\bar n\over (0.39)^2}\hmp2
\eqno(15a)
$$
for quasi-exponential inflation, and
$$
Q_{S}^2 = {\pi^2 \over 10}~p~ f(n) A(\nu)^2 \hmp2
\eqno(15b)
$$
$(\nu - {3\over 2} =  (p - 1)^{-1} = (1 - n)/ 2)~), $
for power law inflation. (See also Fabbri, Lucchin and Mataresse 1986;
Lyth \& Stewart 1992.)

Since the coefficients ${\rm a}_{lm}$ in (3) are Gaussian random variables,
the
variances $Q_{T}^2$ and $Q_{S}^2$ given by (4b) and (15a,b) must be added,
when considering the combined effect of both gravity waves and adiabatic
density perturbations on the anisotropy of the CMBR.
As a result, the net quadrupole anisotropy will be given by
$$
Q_{\rm COBE-DMR}^2 = Q_{S}^2 + Q_{T}^2
\eqno(15c)
$$
where $Q_{\rm COBE-DMR}$ is the {\it rms}-quadrupole normalised amplitude
$Q_{rms-PS} = 5.86\times 10^{-6}$ detected by COBE (Smoot et al. 1992).
{}From (4b) and (15a,b), we find that the ratio $(Q_{T}^2/
Q_{S}^2) = (2.94/p f(n))$ depends sensitively on $p$ -- the exponent in
power law inflation. We find that for $p \le 16$, the
contribution from gravity waves to the CMBR
anisotropy dominates the contribution from scalar density perturbations,
whereas for $p > 16$ the reverse is true. The ratio $(Q_{T}^2 / Q_{S}^2)$
has been plotted against the equation of state of the inflaton field $w$
 $(P = w \rho)$,
in Figure 2 ~($p = (2/ 3(1+w)) > 1$). Substituting the values of
$Q_{T}^2$ and $Q_{S}^2$ in (15c), we obtain
$$
\bigg({H_1\over m_{\rm p}}\bigg)^2 = {Q_{\rm COBE-DMR}^2\over [{\pi^2 \over 10}
{}~p~f(n) + 2.9]A(\nu)^2}\eqno(16b)
$$
for power law inflation. Setting $p=2{\bar N}$ in (16b) we recover the
slow-roll limit of quasi-exponential inflation.
As a result (16b) allows us to determine the Hubble parameter during
inflation both for power law, as well as for quasi-exponential inflation,
(we set ${\bar n} = ({\bar N}/64) = 1$ in this case).
Knowing the value of $( H_1/m_{\rm p})$, we can determine the value of
$Q_S^2$ -- the
fraction of the quadrupole anisotropy contributed
 by density perturbations, as well as $(\delta\rho/\rho)_H^2$ -- the amplitude
of density
 fluctuations at horizon crossing (see (5), (9), (12)).
 We follow this procedure and use the COBE-DMR results
 to predict the value of the {\it rms} mass fluctuation on a given scale

$$\bigg({\Delta M\over M}\bigg)^2(R) = {1\over 2\pi^2} \int{k^2 dk \dk^2
W^2(kR)},
 \eqno(17a)$$

\noindent where $W(kR)$ is the {\it top hat} window function
(Peebles 1980).

Our results for cold dark matter models with power law primordial spectra :
$\dk^2 = A\, k^n~ T_{k,{\rm cdm}}$, (where $T_{k,{\rm cdm}}$ is the CDM
transfer
function given in Appendix~G of Bardeen et al. (1986)~)
are shown in Figure  4, for different values
of the parameters $n$ and $h_{50}$ ($\Omega_{\rm cdm} \simeq
0.9, \Omega_{\rm baryonic} \simeq 0.1$, is assumed).
(Values of n in the range $0.5 \le n \le 1.7$, are consistent
with the COBE - DMR results (~Smoot et al. 1992~).)
For a scale-invariant spectrum $n = 1$,
our results agree with those of Bond and Efstathiou (1984, 1987).
Our results also show that $\dm (16h_{50}^{-1} Mpc) \le 1$ for
$n \le 0.87$. Since mass need not trace light, in models with nonbaryonic
dark matter, our results can be taken to mean that a biasing factor
($b_{16} = (\dm)^{-1}_{16}$) mostly
greater than unity, is required in order to reconcile theoretical models
based on power law inflation , with observations. From Figures  3a,b it follows
that the value of $b_{16}$ is sensitive to both $n$ as well as $h_{50}$,
decreasing with increasing $n$ and $h_{50}$. It may be noted that since
 gravity waves contribute predominantly to the
CMBR anisotropy for $n \le 0.87$, the biasing factor $b_{16h_{50}^{-1}}$
is much larger than it would have been had only the contribution from
scalar density perturbations to the CMBR been considered. This makes CDM
models with power law inflation less compatible with the excess galaxy
clustering observed in the APM survey, than had previously been assumed
(~Maddox et al. 1990; Liddle, Lyth \& Sutherland 1992~).

The value of $(H_1/m_{\rm p})$ obtained from equation (16) also  determines the
amplitude
 of gravity waves generated during inflation, and  normalises the gravity wave
 spectral
energy density -- $\epsilon(\omega) = \omega(d\epsilon_g/ d\omega)~~$.
The ratio of the spectral energy density to the critical energy density
$~\Omega_g(\lambda) = \epsilon(\lambda)/ \epsilon_{cr}~~$, has been plotted
in Figure  4, for gravity waves created during exponential inflation,
quasi-exponential inflation and
power law inflation.
(For the corresponding analytical expressions describing
$\Omega_g(\lambda)$, see Sahni (1990)
and Grishchuk \& Solokhin (1991).)
The amplitude of gravity waves is significantly smaller than the sensitivity
of the current generation of terrestrial bar and beam detectors.
The best hope for the detection of the stochastic gravity wave background
appears to lie with the {\it space - interferometer}, whose development
is still in the conceptual stage (Thorne 1988).

\vskip .4cm
\section  {Acknowledgements}

We benefitted from conversations with Patrick Das Gupta. One of us
(VS) acknowledges stimulating discussions  with Dick Bond. TS  was
financially supported by the Council of Scientific and Industrial
Research, India under its SRF scheme during this work.

\vfill \eject
\section  { References}

\vskip .4cm
\n

\item{}Abbott, L.F. and Wise, M.B., 1984 \np 135B, 279
\vskip .2 cm

\item{}Allen, B., 1988 \pd 37, 2078
\vskip .2 cm

\item{}Bardeen, J.M., Bond, J.R., Kaiser, N, and Szalay, A.S., 1986 \ap 304, 15
\vskip .2 cm

\item{}Bardeen, J.M., Steinhardt, P., and Turner, M., 1983 \pd 28, 679
\vskip .2 cm

\item{}Bond, J.R. and Efstathiou, G., 1984 \ap 285, L45
\vskip .2 cm

\item{}Bond, J.R. and Efstathiou, G., 1987 MNRAS 226, 655
\vskip .2 cm

\item{}Bunch, T. S. and Davies, P. C. W., 1978 Proc. R. Soc. A360, 117
\vskip .2 cm

\item{}Burd, A.B. and Barrow, J.D., 1988 \np B308, 929
\vskip .2 cm

\item{}Fabbri, R., Lucchin, L.F. and Mataresse, S., 1986 \pl 166B, 49
\vskip .2 cm

\item{}Fabbri, R. and Pollock, M.D., 1983 \pl 125B, 445
\vskip .2 cm

\item{}Gradshteyn, I.S. and Ryzhik, M., {\it Table of Integrals, Series and
Products}, (Acad. Press Inc., N.Y., 1980) p. 692
\vskip .2 cm

\item{} Grishchuk, L.P., 1974 Zh. Eksp. Teor. Fiz. 67, 825 [~\jetp 40, 409
(1975)~]
\vskip .2 cm

\item{} Grishchuk, L.P. and Solokhin, M., 1991 \pd 43, 2566
\vskip .2 cm

\item{} Guth, A. and Pi, S.Y., 1982 \prl 49, 1110
\vskip .2 cm

\item{} Halliwell, J.J., 1987 \pl 185B, 341
\vskip .2 cm

\item{} Hawking, S.W., 1982 \pl 115B, 295
\vskip .2 cm

\item{} Hawking, S.W., 1985 \pl 150B, 339
\vskip .2 cm

\item{} Kolb, E.W. and Turner, M.S., {\it The Early Universe},
Addison - Wesley Publishing Company (1990)
\vskip .2 cm

\item{} La, D. \& Steinhardt, P.J., 1989 \prl 62, 376
\vskip .2 cm

\item{} Liddle,  A. R., Lyth, D. H. and Sutherland W. J. 1992, Phys. Lett.
B274, 244

\item{} Linde, A.D., 1982 Phys. Lett. 116B, 335
\vskip .2 cm

\item{} Lucchin, L.F. and Mataresse, S., 1985a \pl 164B, 282
\vskip .2 cm

\item{} Lucchin, L.F. and Mataresse, S., 1985b \pd D32, 1316
\vskip .2 cm

\item{} Lyth, D.H. and Stewart, E.D. 1992 \pl 274B, 168
\vskip .2 cm

\item{} Maddox, S. J., Efstathiou, G., Sutherland, W. J. and
 Loveday J. 1990, MNRAS 242, 43

\item{} Olive, K.A., 1990 Phys. Rep., 190, 309
\vskip .2 cm

\item{} Peebles, P.J.E., {\it The Large Scale Structure of the Universe}
(Princeton University, Princeton (1980)
\vskip .2 cm

\item{} Rubakov, V.A., Sazhin, M.V. and Veryaskin, A.V., 1982 \pl 115B, 189
\vskip .2 cm

\item{} Sachs, R.K. and Wolfe, A.M., 1967 \ap 147, 73
\vskip .2 cm

\item{} Sahni, V., 1988, Class. Quantum Grav. 5, L113

\item{} Sahni, V., 1990 \pd 42, 453
\vskip .2 cm

\item{} Smoot, C. F.  et al. 1992, Astrophys. J. Lett. ({\it Submitted})

\item{} Starobinsky, A.A., 1979 \jetpl 30, 719
\vskip .2 cm

\item{} Starobinsky, A.A., 1982 \pl 117B, 175
\vskip .2 cm

\item{} Thorne, K.S., in {\it 300 Years of Gravitation}, eds. S.W. Hawking
and W. Israel, (Cambridge Univ. Press, Cambridge, 1988) p.330

\item{} Vilenkin, A. and Ford, L. H., 1982 Phys.  Rev. D25, 1231
\vskip .2 cm

\bigskip
\vfil\eject

\section { Figure Captions}

\leftline {\bf Fig. 1}
\vskip .2 cm
The superadiabatic amplification of gravity waves is shown for a ``potential
barrier" $V[a(\eta)] = (\ddot a/ a)$, for inflation $~(a = (\e_o/\e)~)$
followed
by radiation and matter domination $~~(a = a_o \e ( \e + \tilde \e_o)~)$
(~Sahni 1990~).
At early times $t < t_1$,  the scalar field is in its vacuum state $\tilde
\phi_k^{(+)}$.
At late times $t \approx t_0 $, the scalar field will in general be described
by a
linear superposition of positive and negative frequency solutions of eq. (1) :
$\tilde \phi_k (\e) = \alpha \tilde \phi_k^{(+)} + \beta \tilde \phi_k^{(-)}.$
A mode with comoving wavenumber $k = (2\pi a/\lambda)$ is  shown to leave
the Hubble radius during inflation ($t_1$),
and re-enter it during matter domination ($t_0$). (Figure not drawn to scale.)

\bigskip
\leftline {\bf Fig. 2}
\vskip .2 cm
The ratio of the Quadrupole anisotropy from tensor and scalar waves is plotted
against both the equation of state $w$ and the spectral index $n$~ ($\dk^2
\propto k^n$) for power law inflation; $n = (7 + 9w)/ (1 +3w)$.
The maximum value $n = 0.985$ corresponds to $p = 134$ (in $a \propto t^p$),
which is equivalent to $\sim 67$ e-foldings of quasi-exponential inflation.
\bigskip
\leftline {\bf Fig. 3}
\vskip .2 cm
$a.~$ The {\it rms} density contrast on scales
$16 h_{50}^{-1} Mpc - (\Delta M / M)_{16}$, is shown for different values
of the Hubble parameter $ H_o$~ ($= h_{50} \times 50 km sec^{-1} Mpc^{-1}$),
and the spectral index $n~$ (~($n = 1$) corresponds to the Harrison - Zeldovich
spectrum). Some values of the biasing factor $b_{16} =  (\Delta M /
M)^{-1}_{16}$,
are also shown.

\vskip .2 cm

$b.~$ The biasing factor $b_{16} =  (\dm)^{-1}_{16}$, is shown plotted against
the
spectral index $n$, for three values of the Hubble parameter:
$H = $ 50, 75 \& 100 $ km sec^{-1} Mpc^{-1}$. The dotted lines correspond to
$b_{16} = 1$  and $b_{16} = 2.5 $.

\bigskip

\leftline {\bf Fig. 4}
\vskip .2 cm
The spectral energy density of gravity waves (in units of the critical density)
is shown as a function of the wavelength, for exponential inflation,
quasi-exponential inflation (dotted line), and power law inflation $ a \propto
t^p$
 with $p = 21$ and $p = 9$. For comparison, the expected
sensitivity of the Laser Interferometer Gravitywave Observatory (LIGO),
and of the projected ``Beam in space" (space - interferometer)
has also been plotted (Thorne 1988).


\noindent

\vfill\eject
\bye